\documentclass[aps,prl,showpacs,twocolumn,groupedaddress,superscriptaddress]{revtex4}
\usepackage{amsmath,amssymb}
\usepackage{graphicx}
\usepackage{psfrag,color}

\renewcommand\mark[1]{\bgroup #1\egroup}

\begin{document}

\title{Cooperative crosslink (un)binding in slowly driven bundles of semiflexible
  filaments}

\author{Claus Heussinger} \affiliation{Institute for Theoretical Physics,
  Georg-August University of G\"ottingen, Friedrich-Hund Platz 1, 37077
  G\"ottingen}\affiliation{Max-Planck Institute for Dynamics and
  Self-Organization, Bunsenstr 10, 37073 G\"ottingen}

\begin{abstract}
  Combining simulations and theory I study the interplay between bundle elastic
  degrees of freedom and crosslink binding propensity.  
  By slowly driving bundles into a deformed configuration, and depending on the
  mechanical stiffness of the crosslinking agent, the binding affinity is shown
  to display
  a sudden and discontinuous drop. This indicates a cooperative unbinding
  process that involves the crossing of a free-energy barrier.  Choosing the
  proper crosslinker, therefore, not only allows to change the composite elastic
  properties of the bundle, but also the relevant time-scales, which can be
  tuned from the single crosslink binding rate to the much longer escape time
  over the free-energy barrier.
\end{abstract}

 \date{\today}

\maketitle

Filamentous biopolymers, like F-actin or microtubules, play key roles in many
cellular and extra-cellular processes. A variety of different crosslinking
proteins are used in order to assemble these filaments into higher order
structures, like bundles or networks.
Permanently crosslinked reconstituted F-actin networks have been used as simple
model systems to advance the understanding of the mechanics of the more complex
physiological systems~\cite{bausch06,LielegSoftMatter2010,KaszaCOCB2007}.
Similarly, experiments with actin bundles allowed to assess the influence of the
crosslink stiffness on the bundle mechanical
properties~\cite{claessens06NatMat}. A theoretical description was
developed that characterizes bundle mechanics starting from the assumption that
crosslinks are permanently bound~\cite{batheBPJ2008}.
Recent experiments~\cite{lielegPRL2008Transient,broederszArxiv2010} indicate,
however, that low frequency rheology of actin networks can only be understood by
properly accounting for crosslink binding and unbinding
processes~\cite{wolffArxiv2010}.
This points the way towards investigating the possible coupling between filament
elasticity and crosslink
binding.
Several theoretical studies dealt with the equilibrium phase behavior of
crosslinked networks~\cite{zil03}. 
Much less is known for the case of filament bundles.
Refs.~\cite{benetatosPRE2003,kierfeld05PRL} described an equilibrium unbundling
transition in terms of a competition between the entropy of filament bending
fluctuations and the attractive force mediated by crosslink binding.  Similarly,
Ref.~\cite{grasonPRL2007} invoked filament twist elasticity to explain the
apparent thermodynamic stability of bundles with a characteristic
radius~\cite{pelletier03,claessensPNAS2008}.

Going beyond unconstrained thermal equilibrium, I study here the nonlinear
response of a reversibly crosslinked filament bundle to a driving force or
deformation.
In the crowded environment of a cell F-actin bundles are likely to be constantly
exposed to external forces and obstacles that bend, compress or otherwise deform
it.  Little is known about how the bundle, and in particular the crosslinks
inside the bundle, react to such a perturbation.
I will show that bundle deformation leads to crosslink unbinding processes that
crucially depend on the stiffness of the crosslinking protein. In particular,
bundles with stiff crosslinks display a cooperative, discontinuous transition
from a highly to a weakly crosslinked state.

\begin{figure}[ht]
\begin{center}
\includegraphics[clip=,height=0.7\columnwidth]{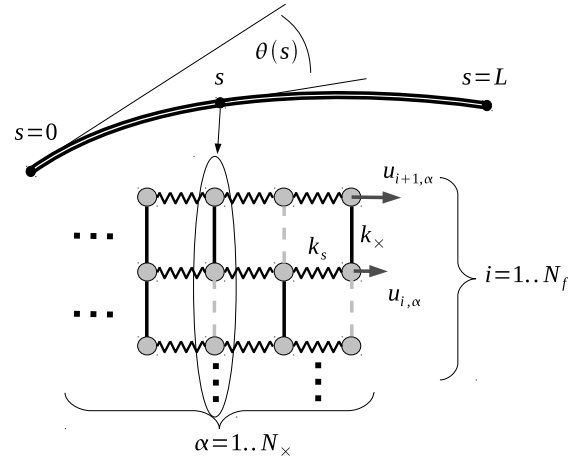}
\caption{Illustration of the bundle geometry. The bundle is bent by an
  arclength-dependent \mark{tangent} angle $\theta(s)$.  Filaments are discretized into
  beads, which can move along the bundle axis (as quantified by
  the displacement $u_{i\alpha}$) but not in lateral direction. In addition,
  each bead can host a crosslink that connects it to its neighboring site on the adjacent
  filament (vertical solid lines: bound crosslink; dashed lines: unbound
  crosslink).
  \label{illustration}}
\end{center}
\end{figure}

Consider a bundle with $N_f$ filaments in two spatial dimensions.  The bundle is
slowly driven into a bent state characterized by the \mark{tangent} angle $\theta(s)$
at arclength position $s=0\ldots L$ along the bundle backbone (see
Fig.~\ref{illustration}). The driving is assumed slow enough such that the
crosslink binding degrees of freedom can equilibrate under the constraint of the
given bending amplitude.

\begin{figure*}[ht]
\begin{center}
\includegraphics[clip=,height=0.43\columnwidth]{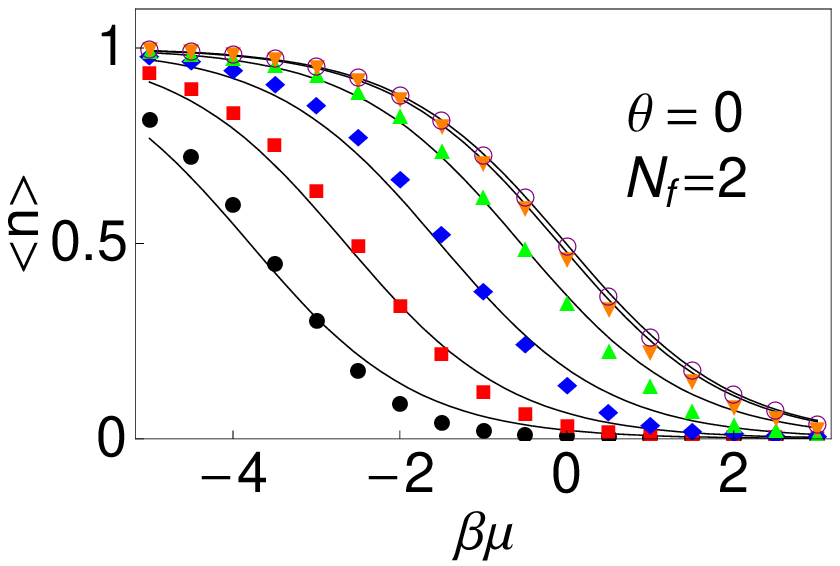}
\includegraphics[clip=,height=0.43\columnwidth]{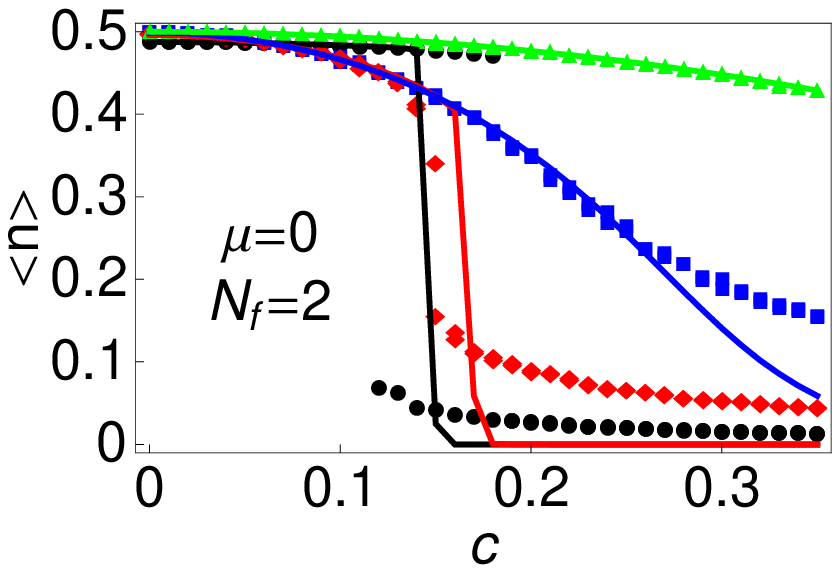}
\includegraphics[clip=,height=0.43\columnwidth]{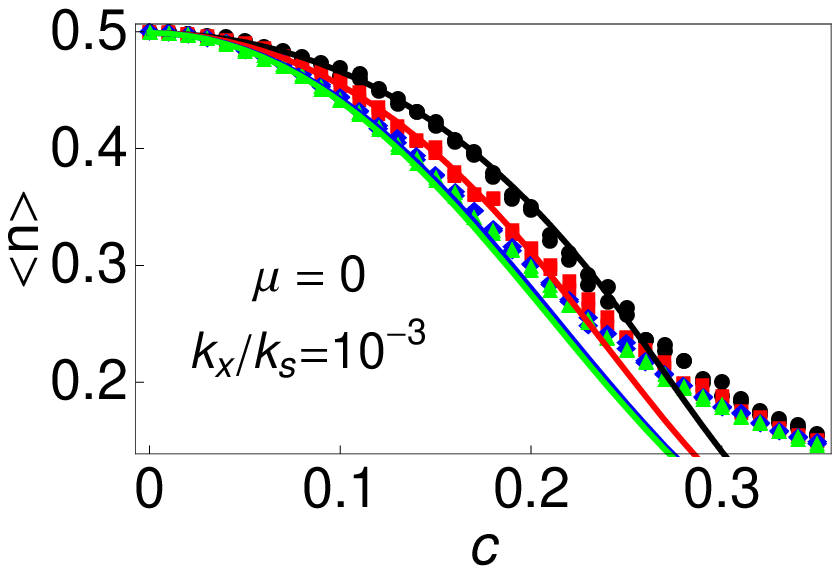}
\caption{ Average crosslink occupation as function of the control parameters
  $\mu$, $k_\times$, $c$ and $N_f$ using $N_\times=100$ and $\beta k_sb^2=100$.
  (left) Undeformed bundle ($\theta\equiv 0$) for different crosslink stiffness
  $k_\times/k_s=10^{-2},\ldots, 10^3$ (from right to left).  The solid lines are
  the solutions of the one-crosslink model, Eq.~(\ref{eq:1crosslink}).
  (center,right) As function of bundle deformation for different crosslink
  stiffness, $k_\times/k_s =10^{-5}\ldots 10^{-2}$ (center) and for different
  filament numbers $N_f=2\ldots 8$ (right). For the largest crosslink stiffness
  (black circles, center) the simulations display considerable hysteresis,
  indicating a strongly first-order transition, where a free-energy barrier does
  not permit to sample efficiently. The lines are obtained without fit
  parameters from the theoretical model,
  Eq.~(\ref{eq:heff}).\label{n.vs.curv}\label{curv.zero}}
\end{center}
\end{figure*}

As a consequence of bundle bending, filaments have to slip relative to each
other, thus bringing the crosslinking sites out of registry and leading to
crosslink deformation. Denote by $u_{i\alpha}$ the axial slip motion of filament
$i$ at arclength position $s_\alpha$ (Fig.~\ref{illustration}), then the
crosslink (shear) deformation can be taken as
$\Delta_{i\alpha}=u_{i+1\alpha}-u_{i\alpha}+b\theta(s_\alpha)$, where $b$ is the
lateral distance between the two filaments that are connected by the crosslink.
In the continuum limit this reduces to the well known expression for the shear
strain, \mark{$\partial u_x/\partial y +\partial u_y/\partial x$}, with the
axial ($u_x$) and the lateral $(u_y\sim \int_s\theta)$ components of the
displacement field. Bringing the two crosslinking sites back into registry
($\Delta\to0$) is therefore only possible if one of the filaments stretches out
farther than its connected partner,
$u_{i+1\alpha}=u_{i\alpha}+b\theta(s_\alpha)$, in order to compensate for the
bending induced mismatch, $b\theta$.  This leads to the competition between
filament stretching and crosslink shearing as a fundamental mechanism governing
bundle mechanics~\cite{heussingerWLB2007}\mark{: Filament stretching can only be
  relieved at the cost of crosslink shearing; alternatively, if crosslinks
  remain unsheared, the filaments automatically have to stretch out.}

By discretizing the filaments at the sites of the crosslinks one can write the
stretching and the shearing energy as a sum over all filaments and all
crosslinking sites as 
\begin{eqnarray}\label{eq:hs}
  H_s & = & \frac{k_s}{2}\sum_{i=1}^{N_f}\sum_{\alpha=1}^{N_\times} \left (u_{i\alpha+1}-u_{i\alpha}  \right
)^2 \\\label{eq:hx}
H_\times & = & \frac{k_\times}{2}\sum_{i=1}^{N_f-1}\sum_{\alpha=1}^{N_\times} n_{i\alpha} \left ( u_{i+1\alpha}-u_{i\alpha}
   +b\theta_\alpha\right )^2
\end{eqnarray}
Filaments are thus characterized by a stretching stiffness, $k_s$ that
constrains the axial motion of two successive sites along a given filament $i$.
Crosslinks with stiffness $k_\times$ couple two filaments laterally, but at the
same arclength position, $\alpha$.  \mark{This energy expression, which has been
  given previously~\cite{batheBPJ2008,heussingerWLB2007}, simply corresponds to
  a discretized version of anisotropic continuum elasticity~\cite{love}.}
The \mark{important new} ingredient is the possibility of crosslink (un)binding,
which is accounted for by introducing the occupation variables,
$n_{i\alpha}=0,1$, corresponding to the unbound and the bound state,
respectively.

Assuming $\theta(s)$ to be constant on the time-scales of interest, the bending
energy of the bundle needs not be considered explicitly, as it contributes an
unimportant constant.
The degrees of freedom of the bundle are then the axial displacements
$u_{i\alpha}$ of the sites, as well as the occupation with crosslinks,
$n_{i\alpha}$. I neglect the possibility of individual filaments performing
lateral motion transverse to the bundle axis. It has been
shown~\cite{benetatosPRE2003,kierfeld05PRL} that at sufficiently low crosslink
concentration the entropy gained by these fluctuations can drive an unbundling
transition.  Being primarily interested in highly crosslinked bundles away from
the unbundling transition, these degrees of freedom can be assumed to be frozen
out~\footnote{Short wavelength bending fluctuations can also be viewed as being
  incorporated into an effective stretching stiffness $k_s^{\rm eff}$ which then
  contains both enthalpic and entropic effects}.

This ends the model definition.  We will now proceed to calculate the average
crosslink occupation $\langle n\rangle$ as a function of the control parameters.
Combining Metropolis Monte-Carlo simulations with theoretical analysis it will
be possible to elucidate the complex interrelationship between crosslink binding
and bundle elasticity.

For the simulations, units are chosen such that $b=k_s=1$.  \mark{The binding
  variables $n_{i\alpha}$ enter Eq.~(\ref{eq:hx}) linearly, so they can be
  traced out exactly~\cite{kierfeld05PRL}. As a result one obtains an effective
  crosslink potential that only depends on the axial displacements
  $u_{i\alpha}$}
\begin{eqnarray}\label{eq:}
  V_\times^{\rm eff} = -k_BT\sum_{i\alpha}\log\left(
    1+e^{-\beta\mu}\exp(-\frac{\beta k_\times}{2}\Delta_{i\alpha}^2)
  \right)\,. 
\end{eqnarray}
The depth of the potential is given by $\beta V(0) =
\log(1+e^{-\beta\mu})$ and corresponds to the standard free energy of a single
crosslink with chemical potential $\mu$.
The potential range is set by the crosslink stiffness, $\Delta_{\rm max} \sim
1/\sqrt{\beta k_\times}$.

Fig.~\ref{curv.zero}a displays the resulting crosslink occupation $\langle
n\rangle$ for an undeformed bundle ($\theta\equiv 0$) as a function of $\mu$ and
$k_\times$.  It comes as no surprise that crosslinks tend to unbind by either
reducing the potential depth ($\mu \nearrow$) or decreasing its range ($k_\times
\nearrow$). Accordingly, already a simple toy-bundle with only one crosslink and
two $u$-degrees of freedom allows to reproduce the simulated curves fairly well.
The Hamiltonian of this simple model reads
\begin{eqnarray}\label{eq:H1dof}
H = \frac{k_s}{2}\left ( u_1^2+u_2^2\right) + \frac{nk_\times}{2}(u_1-u_2+b\theta)^2 + \mu n\,,
\end{eqnarray}
and one obtains for the crosslink occupation ($\theta=0$)
\begin{eqnarray}\label{eq:1crosslink}
\langle n \rangle = \left (1+e^{\beta\mu}\sqrt{1+\frac{2k_\times}{k_s}}\right )^{-1}
\end{eqnarray}
which is shown together with the Monte-Carlo data in Fig.~\ref{curv.zero}a.
The decreasing crosslink occupation is driven by the interplay of binding
energy, as characterized by $\mu$, and the configurational entropy stored in the
axial displacement modes. Notably, there is no sudden unbundling transition as
in the models of Refs.~\cite{benetatosPRE2003,kierfeld05PRL}. In those systems,
it is the entropy stored in the \emph{transverse} bending fluctuations that drives the
transition towards the unbundled state.

Let us now discuss the crosslink occupation as a function of bundle deformation.
For specifity, I take \mark{$\theta(s) = cN_\times\sin(qs)$} with $q=\pi/2L$,
corresponding to an approximate solution for a clamped bundle that is loaded at
its free end. The parameter $c$ reflects the curvature of the bundle, $c\sim
\theta'$, and will serve in the following as a measure of the amplitude of the
imposed deformation.

As is evident from Fig.~\ref{n.vs.curv}b,c the crosslink occupation $\langle
n\rangle$ decreases upon increasing bundle deformation $c$. This agrees well
with the above mentioned notion of bending-induced mismatch between the binding
sites. An increasing mismatch implies increasing strain on the crosslink, which,
in turn, increases their tendency to unbind. Unlike the case discussed in
Fig.~\ref{n.vs.curv}a, however, the decrease of $\langle n\rangle$ is gradual
and smooth \emph{only} below a certain value of the crosslink stiffness
$k_\times$.  Above this value, $\langle n\rangle$ displays a very sudden and
discontinuous drop with bundle deformation. Interestingly, this effect cannot be
explained on the basis of the one-crosslink toy-model of Eq.~(\ref{eq:H1dof}),
which only gives a smooth decay governed by a renormalized chemical potential
\begin{eqnarray}\label{eq:tildemu.1dof}
\tilde\mu = \mu +\frac{(b\theta)^2}{2}\cdot\frac{k_s/2}{1+k_s/2k_\times}
\end{eqnarray}
One concludes, that the discontinuous transition represents a truly cooperative
effect, where many crosslink unbinding events are needed to drive the bundle
into a weakly crosslinked state.

Analytical progress can be made in a mean-field approximation. To this end we go
back to the original Hamiltonian of Eq.~(\ref{eq:hx}) and substitute the actual
crosslink state $n_{i\alpha}$ by its bundle averaged value,
$n=\sum_{i\alpha}n_{i\alpha}/(N_f-1)N_\times$. As compared to the fully occupied
bundle ($n=1$), this has the effect of renormalizing the crosslink stiffness
$k_\times \to nk_\times$.

The advantage of this approximation is that the $u$-variables can now be
integrated out by going to the continuum limit and transforming to Fourier
space. The resulting effective free energy (per crosslink), which still depends
on the bundle averaged occupation variable $n$ has the rather simple
structure~\footnote{As for the MC simulations I take $\theta(s) \sim \sin(qs)$
  and consider the lowest mode only, $q=\pi/2L$.}
\begin{eqnarray}\label{eq:heff}
  F_{\rm eff}(n) &=& \frac{A}{1+n_0/n} + \mu n \\\nonumber
  &+& k_BT\left[n\log(n) +(1-n)\log(1-n)\right]\,,
\end{eqnarray}
Next to the chemical potential $\mu$, there are two relevant parameters. The
crosslink stiffness is encoded in $n_0 \simeq
(k_s/k_\times)N_f(N_f+1)/N_\times^2$, while the parameter $A$ is related to the
amplitude of the bending deformation, $A\simeq k_sb^2c^2N_f(N_f+1)$. The term with the
logarithm corresponds to the standard entropy of mixing and counts the number of
states that are compatible with a given average crosslink occupation.
Calculating the average $\langle n\rangle$ from Eq.~(\ref{eq:heff}) can easily
be performed numerically, and is plotted as solid lines in
Fig.\ref{n.vs.curv}b,c.  As can be seen the model and the simulation are in good
agreement without fit parameters.

At large bending amplitudes some deviations are apparent.
By looking more closely into the distribution of crosslinks inside the bundle
one finds that this state is actually highly inhomogeneous, with many more
crosslinks bound close to the clamped end than at the free end.  In fact, the
discontinuous reduction of bound crosslinks proceeds (with our MC-''dynamics'')
via an unzipping mechanism, where the crosslinks at the free end unbind first.
An unbinding front then moves rapidly towards the clamped end of the bundle
where it stabilizes some distance away.  \mark{A similar inhomogeneity can occur
  in the lateral direction, with crosslinks on different filaments unbinding at
  different bending amplitudes, leading to a sequence of discontinuous
  transitions, each corresponding to the unbinding of a single filament pair.
  I leave the detailed discussion of these complex
  unbinding pathways for a future publication.}

\begin{figure}[ht]
  \begin{center}
    \includegraphics[clip=,height=0.5\columnwidth]{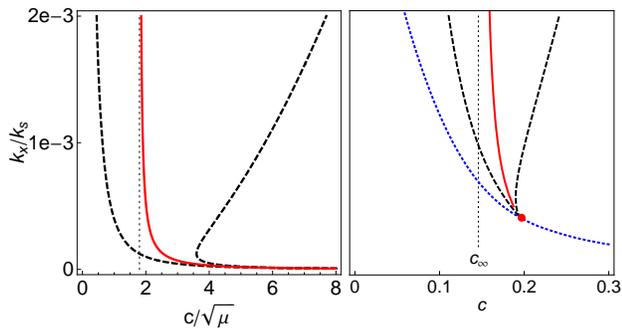}
    \caption{Phase-diagram from a saddle-point analysis of Eq.~(\ref{eq:heff})
      at $T=0$ (left) and for finite temperatures (right). In the latter case,
      the discontinuous transition (solid red line) terminates at a critical
      point, which shifts with the chemical potential along the (dotted) blue
      line. Metastable states exist within region delineated by (dashed) black
      lines. \mark{In the limit of large crosslink stiffness $k_\times\to\infty$
        (thin vertical line) the discontinuous transition asymptotes at
        $c_\infty\sim 1/\sqrt{N_f(N_f+1)}$.}\label{phasediagram}}
\end{center}
\end{figure}

More insight into the physical mechanism that leads to the observed
discontinuous transition can be obtained with the help of a saddle-point
analysis. By minimizing Eq.~(\ref{eq:heff}) with respect to $n$, $dF_{\rm
  eff}/dn=0$, one obtains the phase diagram depicted in Fig.~\ref{phasediagram}.
At $T=0$ (left figure), i.e. without the entrop of mixing, the model has two
minima at $\langle n\rangle=0,1$ that coexist in a certain range of
parameter
values (limited by the dashed black lines). The condition $F_{\rm eff}(0) { =}
F_{\rm eff}(1)$ then gives a line of discontinuous transitions (solid red line)
\mark{where the crosslink occupation jumps from $\langle n\rangle=1$ to $\langle
  n\rangle=0$.  For large crosslink stiffness, the bending amplitude at the
  transition depends on the number of filaments as $c_\infty\simeq
  (N_f(N_f+1))^{-1/2}$ (vertical dashed line).}

At finite temperatures (right figure) this line terminates at a critical point,
beyond which the crosslink occupation decreases smoothly with \mark{increasing
  bending amplitude}. The critical point itself depends on the chemical
potential $\mu$ as indicated in the figure by the (dotted) blue line.  Thus,
stiff crosslinks
tend to unbind cooperatively in a discontinuous transition, while soft
crosslinks
unbind one after the other leading to a smooth decrease of the average crosslink
occupation.

This qualitative difference between bundles with soft or stiff crosslinks can be
intuitively understood as follows. When the crosslinks are soft, they cannot
induce any stretching in the filaments.  Accordingly the force in the crosslinks
is small and set by the crosslink stiffness $k_\times$. In this limit unbinding
events are purely local with no consequences for the rest of the bundle. In
contrast, stiff crosslinks are able to drive the filaments into a stretched
configuration.  The force in the crosslink is now set by the filament stretching
stiffness $k_s$. An unbinding event will then affect the force balance in the
filaments, with the potential of influencing the bundle state far away from the
unbinding site. Note, that a similar distinction between soft and stiff
crosslinks has been observed in Ref.~\cite{PhysRevLett.103.238102}, dealing with
the phase behavior of helical filament bundles.

The discontinuous transition and the associated hysteresis observed in the
simulations (see Fig.\ref{curv.zero})b) is tantamount to a long time-scale
associated with the escape over a free-energy barrier that separates highly
crosslinked from weakly crosslinked states. Thus, the choice of crosslinker not
only determines the mechanical properties of a bundle but also affects the
relevant relaxation times. Depending on the stiffness of the crosslink, this
time-scale may range from the single crosslink binding rate to the escape time
over the free-energy barrier.

Such a long time-scale may find its analogy in a recent experiment with F-actin
bundles crosslinked by $\alpha$-actinin~\cite{strehleEBPJ2011}.  In the
experiment a bundle was released from a bent configuration after keeping it
there for a certain waiting time. If the waiting time was long enough, the
bundle did not relax back into the expected straight state, but remained with a
considerable residual bending deformation. It seems that under bundle
deformation new binding sites become available that stabilize the bent shape by
allowing the crosslinks to rebind in a ``native'', unstrained state.
Interestingly, the apparent time-scale (the waiting time) necessary to observe
residual bundle bending was considerably longer than the time required for
(un)binding of the individual $\alpha$-actinin linker.
By taking $\mu=0$ one can set up a simple correspondence between unbinding
events as defined in the present work, and rebinding events into the new sites
in the experiment. With this mapping the long time-scale observed in the
experiment can be identified with the escape time over the free energy barrier,
\mark{which is estimated as $\tau\approx 2^{N_{\rm tot}}/N_{\rm tot}$, where
  $N_{\rm tot}=(N_f-1)N_\times$ is the total number of binding
  sites~\cite{tbp,erdmannJCP2004}.}  In this spirit, one would call
$\alpha$-actinin, the crosslinker used in the experiment, a stiff crosslink. It
would be interesting to repeat the experiment with a softer crosslink. Future
experiments should furthermore measure the force needed to bend the bundle, as
this is expected to scale with crosslink occupation and \mark{ display a
  discontinuous drop, $\Delta F\sim N_f^2$~\cite{tbp}.} In particular, it would
be interesting to discuss rate-dependent bundle bending.  This would be a
natural next step in the effort to understand the full dynamics of the coupled
processes of bundle conformation and crosslink binding.  \mark{Finally, it
  should be pointed out that crosslinked actin networks typically consist of
  composite mixtures of single filaments and small
  bundles~\cite{LielegSoftMatter2010}. Therefore, the observed unbinding
  transition could directly be relevant for F-actin network rheological
  properties.}

\acknowledgments

Fruitful discussions with M. Bathe, E. Frey, D. Strehle and R. Vink are
gratefully acknowledged.


\end{document}